\documentclass[12pt]{article}

\usepackage{amsmath,amsthm,amssymb,color,undertilde}
\usepackage{epsfig,amsfonts,latexsym, hyperref, bbm, mathrsfs}
\newcommand{\beq}{\begin{equation}}
\newcommand{\eeq}{\end{equation}}
\title{Particle on a Torus Knot:\\ Constrained Dynamics and Semi-Classical Quantization in a Magnetic Field}

\begin{document}
	
	{\bf \maketitle}
	
\begin{center}
	Praloy DAS \footnote{E-mail: praloydasdurgapur@gmail.com}, Souvik PRAMANIK \footnote{E-mail:  souvick.in@gmail.com} and
	Subir GHOSH \footnote{E-mail: subirghosh20@gmail.com }\\
	\vspace{0.3 cm}
	\small{\emph{Physics and Applied Mathematics Unit, Indian Statistical Institute\\
			203 B. T. Road, Kolkata 700108, India}} \\
\end{center}

\begin{abstract}
	Kinematics and dynamics of a particle moving on a torus knot poses an interesting problem as a constrained system. In the first part of the paper we have derived the modified symplectic structure or Dirac brackets of the above model in Dirac's Hamiltonian framework, both in toroidal and Cartesian coordinate systems. This algebra has been used to study the dynamics, in particular small fluctuations in motion around a specific torus. The spatial symmetries of the system have also been studied.

	In the second part of the  paper we have considered the quantum theory of a charge moving in a torus knot in the presence of a uniform magnetic field along the axis of the torus in a semiclassical quantization framework. We exploit the  Einstein - Brillouin - Keller (EBK)  scheme of  quantization that is appropriate for  multidimensional
	systems.   Embedding of the knot on a specific torus is inherently two dimensional that gives rise to two quantization conditions.  This shows that although the system, after imposing the knot condition  reduces to a one dimensional system, even then it has manifest non-planar features which shows up again in the study of  fractional angular momentum.  Finally we compare the results obtained from  EBK (multi-dimensional) and  Bohr-Sommerfeld (single dimensional) schemes.   The energy levels and fractional spin depend on the torus knot parameters that specifies its non-planar features. 	Interestingly, we show that there can be non-planar corrections to the planar anyon-like fractional spin.
\end{abstract}	


\section{Introduction}
Hamiltonian analysis of constraint systems, as formulated by \cite{dir}, plays an important role in quantization of constraint systems. In this framework, for a special type of constraints, known as Second Class constraints \cite{dir} one needs to replace the canonical Poisson brackets (or symplectic structure) by a new form of brackets, known as Dirac brackets. Classical analysis with Dirac brackets generates the dynamics that is consistent with the constraints. To quantize such a system the Dirac brackets are elevated to quantum commutation relations with $i\hbar$ factor.  In general, the Dirac brackets can be much more complicated that the Poisson brackets and there can appear many debatable issues (such as operator ordering problems, inequivalent quantization, ...(see \cite{kli} for a specific example)) in a straightforward quantization program.

 The Poisson brackets for a system are 
 \beq~~\{x_i,p_j\}=\delta_{ij} ,\{x_i,x_j\}=0,\{p_i,p_j\}=0.
 \label{sp}
 \eeq
 In general, depending on the constraints operating in a particular system, additional terms on the RHS appear in Dirac brackets. It becomes specially interesting  if the  additional terms are non-constant and $x_i,p_i$-dependent, for which the dynamics becomes qualitatively different.

  In the present work we will study models where particles are allowed to move on a restricted form of configuration space which in turn generates Dirac brackets having coordinate dependent additional terms in the Dirac brackets. A prototype of this is the well studied motion of particle in a circle and its quantization. In this paper, we will consider a next level of complication where a particle follows the trajectory of a   {\it{torus  knot}}, that is, its path is a closed loop with a knot embedded on a torus.

  Quantum dynamics of a particle moving on a torus knot  has been studied recently in \cite{sree}.  As a recent application in a very distinct area, we mention that in the context of cosmology, in \cite{torus} particle motion on a torus (without the knot) has been considered.  
 
 Broadly the paper is divided into two parts. In the first part we deal with formal aspects of the problem of particle moving on a torus as a constrained system.  In the present paper we treat these particle models  as constrained systems and Dirac's Hamiltonian analysis of constraint systems \cite{dir} provides a   unified framework for distinct types of restricted particle motion. Indeed the constraints induce a non-canonical phase space structure that can be identified with the Dirac bracket algebra, that was introduced by Dirac to replace the canonical Poisson bracket algebra. The effect of this change in brackets is directly manifested in the dynamics and symmetry transformations that reveal the kinematics. For the particle moving on a torus knot, the constraint analysis appears to be quite involved with a unique feature of the respective Dirac brackets: there does not appear to be a smooth limiting procedure   to reduce the Dirac algebra to Poisson algebra.  The probable reason for this is the topological nature of the constraints involved. Indeed it will be very interesting if these closed paths reveal the presence of non-trivial holonomies.

 The semiclassical quantization of a charged particle on a torus knot in a magnetic field poses an interesting problem. On the one hand, in toroidal coordinate  the system is effectively one dimensional which should require a single quantization condition. But, on the other hand, embedding the knotted path on a torus is inherently three dimensional that reduces to a two degrees of freedom problem for a fixed torus. We perform a semi-classical quantization along   Einstein - Brillouin - Keller (EBK)  scheme    that is appropriate for  multidimensional
 systems.
 
   In three (space) dimensions particle excitations come with integer or half integer spin obeying Bose or Fermi statistics, respectively. Anyon excitations with arbitrary fractional spin and subsequent fractional statistics,  proposed by Wilczek \cite{wil}, occur in non-relativistic planar physics.  The simplest fractional spin model in non-relativistic regime consists of a charge moving in a circle around a solenoid with a magnetic field $B$ perpendicular to the plane of motion. The question is that is it possible to generalize this phenomenon in three space dimensions? In the present paper we will study that. Indeed, non-planar effects can affect the properties of anyons in interesting and non-trivial ways. The quantized energy levels and spin are modified by some terms, that explicitly depend on non-planar parameters of motion. As an example we   generalize the path of the charge (from a planar circle as above) to a path in $R_3$ in the form of  a torus knot  in the presence of a uniform $B$ along the axis of the toroidal path.

The paper is organized as follows: In Section 2, we describe the Hamiltonian constraint analysis as formulated by Dirac \cite{dir}. Furthermore, we explain our problem of interest: particle on a torus knot. Indeed, it consists of two subsections, the first one uses a toroidal coordinate system whereas the second one uses Cartesian coordinate system. Apart from the dynamics we have studied the symmetry properties and small fluctuations of the particle motion about the torus knot. In Section 3, we provide the semi-classical quantization of system of a charged particle on a torus knot in presence of a magnetic field, following EBK quantization program. We conclude in Section 4, where our conclusions as well as future prospects are mentioned.

\section{Particle on a torus knot: classical aspects}
We will formulate the problem both in toroidal and Cartesian coordinate systems since both has specific utilities.
\subsection{Toroidal coordinate system}
The toroidal coordinate coordinate system appears to be the natural choice for the present analysis. The toroidal coordinates are related to the usual cartesian coordinates as below \cite{sree}:
\beq
x_1=\frac{a\sinh\eta\cos\phi}{\cosh\eta-\cos\theta},~~x_2=\frac{a\sinh\eta\sin\phi}{\cosh\eta-\cos\theta},~~x_3=\frac{a\sin\theta}{\cosh\eta-\cos\theta}.
\label{t1}
\eeq
The variables span $0\leq \eta \le \infty ,~-\pi \le \theta \leq \pi,~0 \leq \phi \le 2\pi $. Fixing $\eta $ to eg. $\eta _0$ (as will be done later) indicates a specific toroidal surface and the parameter $a$ and $\eta _0$ are given by $a^2=R^2-d^2,~cosh \eta_0=R/d$ with $R$ and $d$ giving the major and minor radius of the torus respectively.
The inverse transformations are  given by the expressions,
\beq
\eta =\ln\frac{d_1}{d_2},~ \cos\theta=\frac{r^2-a^2}{\left((r^2-a^2)^2+4a^2z^2\right)^\frac{1}{2}},~\phi=\tan^{-1}\frac{x_2}{x_1},
\label{t2}
\eeq
where $$d_1^2=(\sqrt{x_1^2+x_2^2}+a)^2+x_3^2 ~~ , ~~ d_2^2=(\sqrt{x_1^2+x_2^2}-a)^2+x_3^2.$$
The constraint that forces the particle to move in a knot is imposed as $p\theta+q\phi \approx 0$ where $p$ and $q$ are mutually prime numbers. But before proceeding further the torus knot needs to be defined properly \cite{knot}.
Mathematically a knot is a simple, closed, non-self-intersecting curve in $R_3$. According to Knot Theory, a torus knot is a specific kind of knot that can be embedded on the surface of an
un-knotted torus in $R_3$. The $(p,q)$-torus knot winds  $p$ times around the rotational symmetry axis of the torus and $q$ times around a circle in the interior of the torus provided  $p,q$  are  relatively prime. A torus knot is trivial iff  either
$p$ or $q$ is equal to $1$ or $-1$.   Trefoil knot is the simplest
nontrivial example of a  $(2,3)$-torus knot. Operationally a  $(p,q)$-torus knot can be obtained by  identifying  $\theta\rightarrow q\tilde\theta$ and $\phi \rightarrow p\tilde\theta$ in the toroidal surface (\ref{t1}), so that $\tilde\theta = \theta /q=\phi /p $ which is equivalent to the constraint $p\theta -q\phi =0$ indicating a full cycle consists of $\theta \rightarrow \theta + 2\pi q, \phi \rightarrow \phi + 2\pi p$. Note that this parameterization is equivalent to the one used above but for a trivial change of sign.

 For the time being we do not impose any constraint on $\eta $ so that the motion considered here is more general than \cite{sree}{\footnote{We will briefly discuss the motivation later in this section when we study the $\eta$-fluctuations.}. Later on we will fix $\eta =\eta _0$ which actually will be another Hamiltonian constraint that will  further constrain the particle to perform the knot on a  specific torus.

The constrained  Lagrangian for the particle is,
\beq
L=\frac{ma^2(\dot{\eta}^2+\dot{\theta}^2+\sinh^2\eta \dot{\phi}^2)} {2(\cosh\eta-\cos\theta)^2}-\lambda(p\theta+q\phi).
\label{t3}
\eeq
The conjugate momenta are,
\beq
p_\eta=\frac{ma^2\dot{\eta}}{(\cosh\eta-\cos\theta)^2}, ~~
p_\theta=\frac{ma^2\dot{\theta}}{(\cosh\eta-\cos\theta)^2}, ~~
p_\phi=\frac{ma^2\sinh^2\eta\dot{\phi}}{(\cosh\eta-\cos\theta)^2}.
\label{t4}
\eeq
The  Hamiltonian is obtained as,
\begin{eqnarray}
H=\frac{(\cosh\eta-\cos\theta)^2}{2ma^2}\left[p_\eta^2+p_\theta^2+\frac{p_\phi^2}{\sinh^2\eta}\right]+\lambda(p\theta+q\phi).
\label{t5}
\end{eqnarray}

Once again the set of Second Class constraints read,
\begin{eqnarray}
\chi_1=p\theta+q\phi ,~~\dot{\chi_1}\equiv  \chi_2=\frac{(\cosh\eta-\cos\theta)^2}{ma^2}\left[pp_\theta+\frac{qp_\phi}{\sinh^2\eta}\right].
\label{t6}
\end{eqnarray}
As discussed in Appendix II, the Dirac brackets are computed in a straightforward way:
\begin{eqnarray}
&& \lbrace\eta,p_\eta\rbrace=1,  ~  \lbrace\theta,p_\theta\rbrace=\frac{\alpha^2}{\alpha^2+\sinh^2\eta},  ~       \lbrace\phi,p_\phi\rbrace=\frac{\sinh^2\eta}{\alpha^2+\sinh^2\eta} \nonumber \\
&& \lbrace\eta,\theta\rbrace=0, ~    \lbrace\eta,\phi\rbrace=0, ~  \lbrace\eta,p_\theta\rbrace=0,~
\lbrace\eta,p_\phi\rbrace=0,   \lbrace\theta,\phi\rbrace=0,  \lbrace\theta,p_\eta\rbrace=0,\nonumber \\
&& \lbrace\theta,p_\phi\rbrace=\frac{\alpha\sinh^2\eta}{\alpha^2+\sinh^2\eta},~ \lbrace\phi,p_\eta\rbrace=0,~ \lbrace\phi,p_\theta\rbrace=\frac{\alpha}{\alpha^2+\sinh^2\eta},
\nonumber \\
&& \{ p_\eta,p_\theta  \} =-\frac{2\alpha\cosh\eta}{\sinh\eta(\alpha^2+\sinh^2\eta)}p_\phi,~
\lbrace  p_\eta,p_\phi\rbrace=-\frac{2\alpha^2\cosh\eta}{\sinh\eta(\alpha^2+\sinh^2\eta)}p_\phi, \nonumber \\
&& \lbrace p_\theta,p_\phi\rbrace=0,
\label{t7}
\end{eqnarray}
where $\alpha=-\frac{q}{p}$ following \cite{sree} so that $\theta - \alpha \phi =0$. The coordinate $\eta $ behaves in a canonical fashion whereas $p_\eta $ does not. The $\theta $ and $\phi $ sectors behave in a similar way since they are related (by a scaling) by the constraint. Before proceeding further there are a few intriguing aspects of the Dirac bracket structure that is to be noted:\\
(i) the parameter $a$ is absent in the Dirac algebra that depends only on $\alpha $.\\
(ii)interestingly there is no limiting value of $\alpha$ for which the Dirac algebra reduces to the canonical one. From hindsight, we believe that this feature is probably connected to the topologically non-trivial path (torus knot) followed by the particle. This is further manifested by our inability to construct a Darboux like map  that relates  the noncanonical  variables to a set of canonical variables. 

{\bf{Dynamics}}: On the constraint surface the Hamiltonian reduces to
\begin{eqnarray}
H=\frac{(\cosh\eta-\cos\alpha\phi)^2}{2ma^2}\left[p_\eta^2+\frac{p_\phi^2}{\sinh^2\eta}(1+\frac{\alpha^2}{\sinh^2\eta})\right]
\label{t8}
\end{eqnarray}
The equation of motion for $\phi$ is given by,
\begin{eqnarray}
\ddot{\phi} &=& \frac{2(\cosh\eta-\cos\alpha\phi)^3}{m^2a^4\sinh\eta}\left[1-\frac{\cosh\eta(\cosh\eta-\cos\alpha\phi)}{(\alpha^2+\sinh^2\eta)}\right]p_\eta p_\phi \nonumber \\
&& +\alpha\sin\alpha\phi \frac{(\cosh\eta-\cos\alpha\phi)^3}{m^2a^4\sinh^4\eta}p_\phi^2 
-\alpha\sin\alpha\phi  \frac{(\cosh\eta-\cos\alpha\phi)^3}{m^2a^4(\alpha^2+\sinh^2\eta)}p_\eta^2 .
\label{t9}
\end{eqnarray}
The equation of motion for $\eta$ turns out to be
\begin{eqnarray}
\ddot{\eta} &=& \sinh\eta \frac{(\cosh\eta-\cos\alpha\phi)^3}{m^2a^4}~ p_\eta^2+2\alpha\sin\alpha\phi \frac{(\cosh\eta-\cos\alpha\phi)^3}{m^2a^4\sinh^2\eta}~ p_\eta p_\phi \nonumber \\
&& -\frac{(\cosh\eta-\cos\alpha\phi)^3(\alpha^2+\sinh^2\eta)}{m^2a^4\sinh^3\eta}\left[1-\frac{\cosh\eta(\cosh\eta-\cos\alpha\phi)}{\alpha^2+\sinh^2\eta}\right]p_\phi^2.
\label{eeta}
\end{eqnarray}
Replacing the momenta to get the equation fully in configuration space, 
\begin{eqnarray}
\ddot{\phi}&=&\frac{\alpha\sin(\alpha\phi)}{(\cosh\eta-\cos\alpha\phi)}~\dot{\phi}^2-\frac{\alpha\sin(\alpha\phi)}{(\cosh\eta-\cos\alpha\phi)(\alpha^2+\sinh^2\eta)}~\dot{\eta}^2 \nonumber \\
&& +\frac{2\sinh\eta}{(\cosh\eta-\cos\alpha\phi)}\left[1-\frac{\cosh\eta(\cosh\eta-\cos\alpha\phi)}{(\alpha^2+\sinh^2\eta)}\right]\dot{\eta}\dot{\phi},
\label{tt9}
\end{eqnarray}	
\begin{eqnarray}
\ddot{\eta}&=&\frac{\sinh\eta}{(\cosh\eta-\cos\alpha\phi)}~\dot{\eta}^2+\frac{2\alpha\sin\alpha\phi}{(\cosh\eta-\cos\alpha\phi)}~\dot{\eta}\dot{\phi} \nonumber \\
&& -\frac{\sinh\eta(\alpha^2+\sinh^2\eta)}{(\cosh\eta-\cos\alpha\phi)}\left[1-\frac{\cosh\eta(\cosh\eta-\cos\alpha\phi)}{\alpha^2+\sinh^2\eta}\right]\dot{\phi}^2.
\label{eeeta}
\end{eqnarray} 

For a quick check on the consistency of our approach let us impose the other constraint $\eta -\eta_0\approx 0$ which forces the particle to perform it's knotted motion on a fixed torus. The proper way is to start from the beginning and introduce two constraints $\eta -\eta_0\approx 0,~ \chi_1=\theta -\alpha \phi \approx =0$. Demanding time persistence of this set will induce  $p_\eta \approx 0$ and $\chi_2$ respectively. Fortunately the sets $\eta -\eta_0\approx 0,~p_\eta \approx 0$ and $\chi_1,\chi_2 $ mutually commute such that the full $4\times4$ constraint matrix appears in block diagonal form and one construct Dirac Brackets successively using Dirac Brackets from the first set as the starting bracket for the second set. Clearly the set $\eta -\eta_0\approx 0,~p_\eta \approx 0$ does not affect the remaining variables and one is allowed to substitute $\eta =\eta_0$ and $p_\eta=0$ strongly in the algebra. Thus we are left with a system consisting of a single pair of phase space variables with the bracket,
\beq
\lbrace\phi,p_\phi\rbrace=\frac{\sinh^2\eta _0}{\alpha^2+\sinh^2\eta _0},
\label{t10}
\eeq
along with the Hamiltonian
\begin{eqnarray}
H=\frac{(\cosh\eta _0-\cos\alpha\phi)^2}{2ma^2}\frac{1}{\sinh^2\eta _0}\left(1+\frac{\alpha^2}{\sinh^2\eta _0}\right)p_\phi^2.
\label{t11}
\end{eqnarray}
The equation of motion for $\phi $ turns out to be
\begin{eqnarray}
\ddot{\phi}-\alpha \sin(\alpha\phi) \frac{\dot{\phi}^2}{\cosh\eta_0-\cos\alpha\phi} =0,
\label{t12}
\end{eqnarray}
that has been used in \cite{sree} with the solution
\begin{eqnarray}
\tan(\alpha\phi)=\left[\frac{\cosh\eta_0-1}{\cosh\eta_0+1}\right]^{\frac{1}{2}} \tan\left(\frac{A\alpha\sinh\eta_0}{2a}t\right),
\label{sr1}
\end{eqnarray} where $A$ is a constant.

Notice that the R.H.S of the bracket in (\ref{t10}) has become a constant and so the quantization (as has been done in \cite{sree}) can be carried through without any difficulty. Another interesting observation is that
\begin{equation}
\frac{d}{dt}[\sqrt{f}\dot{\phi}]=0
\label{int}
\end{equation}
where $$f(\phi)=\frac{a^2}{(\cosh\eta_0-\cos\alpha\phi)^2}$$ and $$\dot{\phi}=\frac{(\cosh\eta_0-\cos\alpha\phi)^2}{ma^2\sinh^2\eta_0}p_\phi .$$ This also agrees with \cite{sree}.

{\bf{$\eta $-perturbations:}} 
The $\eta$-fluctuations are considered in such a way that the relation between angles $\phi$ and $\theta$ as a constraint is strictly maintained. This means that the torus knot structure of the particle configuration space is deformed without changing the topology. Indeed, it is possible that the particle on a torus can play the role of a toy model in the context of 
 topology induced modification in gauge theory vacuum state \cite{example}, anyonic spin statistics \cite{wil} among others. In these cases the $\eta$-perturbations might be interpreted as some form of "vibrational" excited states.

Next we will study small perturbations of $\eta $ about a constant and large value $\eta_0$. The equations (\ref{tt9}) and (\ref{eeeta}), for large $\eta $, simplify to 
\begin{eqnarray}
	\ddot{\eta} &=& \dot{\eta}^2+2\alpha\frac{\sin\alpha\phi}{\cosh\eta}
	\dot{\eta}\dot{\phi} , \label{eta11} \\
	\ddot{\phi} &=& \alpha\frac{\sin\alpha\phi}{\cosh\eta}(\dot{\phi}^2-\frac{\dot{\eta}^2}{\cosh^2\eta}) .\label{eta12}
\end{eqnarray}
Let us consider small perturbations  $\bar\eta $. To first order we consider solutions of the form $\phi=\phi_s+\overline{\phi}$ and $\eta=\eta_0+\overline{\eta}$ where $\phi_s$ is the known solution (\ref{sr1}) for constant $\eta _0$. We assume a large value for $\eta_0$ and keep only terms of $O(\bar{\eta}),~ O(\bar{\phi})$. This approximation leads to  the equation governing $\bar{\phi},~\bar{\eta}$,
\begin{eqnarray}
	\ddot{\overline{\eta}} &=& \frac{2\alpha\sin(\alpha\phi_s)\dot{\overline{\eta}}\dot{\phi_s}}{\cosh\eta_0}, \\
	\ddot{\overline{\phi}} &=& -\alpha\frac{A^2}{a^2} t \sin\left[\frac{A\alpha\sinh\eta_0}{2a}t\right]\sinh\eta_0 + 2\alpha\frac{A}{a} \sin\left[\frac{A\alpha\sinh\eta_0}{2a}t\right]\dot{\overline{\phi}} + \alpha^2 \frac{A^2}{a^2} \cos\left[\frac{A\alpha\sinh\eta_0}{2a}t\right] \cosh\eta_0~\overline{\phi} \nonumber\\
\end{eqnarray}
where we have used the equation (\ref{sr1}),
$$\tan(\alpha\phi_s(t))=\left[\frac{\cosh\eta_0-1}{\cosh\eta_0+1}\right]^\frac{1}{2} \tan\left(\frac{A\alpha\sinh\eta_0}{2a}t\right)$$ and exploiting  (\ref{int}) we get $$\dot{\phi_s}=\frac{A}{a}(\cosh\eta_0-\cos\alpha\phi_s).$$
In the above  $A$ is a constant. Below we provide the $\bar {\phi }$ equation in large $\eta _0$ limit,
\begin{eqnarray}
	\ddot{\overline{\phi}}=-\alpha\frac{A^2}{a^2}\sin\left[\frac{A\alpha\sinh\eta_0}{2a}t\right]\sinh\eta_0~t+\alpha^2\frac{A^2}{a^2}\cos\left[\frac{A\alpha\sinh\eta_0}{2a}t\right]\cosh\eta_0~\overline{\phi},
\end{eqnarray}
that we have not attempted to solve. On the other hand,
it is straightforward to solve for $\bar{\eta}$,
\begin{eqnarray}
	\overline{\eta}(t)=t-\frac{4a}{\alpha A}\frac{\sin\left[\frac{A\alpha\sinh\eta_0t}{2a}\right]}{\sinh\eta_0\cosh\eta_0}.
	\label{beta}
\end{eqnarray}
The above indicates that the $\eta $-fluctuations will grow linearly with time with a high frequency oscillating behavior impressed upon it.

{\bf{Symmetry properties}:} Let us now analyze the rotation properties of the degrees of freedom. For the unconstrained case the  angular momentum in toroidal coordinates is expressed as   
\begin{eqnarray}
\vec{j} &=& -\frac{\sin\theta\cosh\eta}{\sinh\eta}p_\phi   \hat{\eta}+\cos\theta ~p_\phi\hat{\theta}+[\sin\theta\cosh\eta~ p_\eta-\cos\theta\sinh\eta~ p_\theta]\hat{\phi} \nonumber \\
&=& j_\eta\hat{\eta}+j_\theta\hat{\theta}+j_\phi\hat{\phi},
\label{jj1}
\end{eqnarray}
where the  identities connecting Cartesian and toroidal unit vectors are used. Using the Dirac brackets,  the  transformation of coordinates under rotation are obtained as 
\begin{eqnarray}
\{\vec j,\eta \} =-\sin\theta\cosh\eta ~~\hat{\phi},~~ \{\vec j,\theta \} =\cos\theta\sinh\eta~~\hat{\phi},~~ \{\vec j,\phi \} =\frac{\sin\theta\cosh\eta}{\sinh\eta}~~\hat{\eta}-\cos\theta~~\hat{\theta}.
\label{equation}
\end{eqnarray}
On the other hand, from (\ref{jj1}), for the constrained system, with $\eta =\eta_0,~\theta =\alpha \phi$ as strong equations we find 
\begin{eqnarray}
\vec J&=&-\frac{sin(\alpha\phi)\cosh\eta_0}{\sinh\eta_0}p_\phi~ \hat{\eta} -\frac{\alpha\cos(\alpha\phi)}{\sinh\eta_0}p_\phi~\hat{\theta} +\cos(\alpha\phi) p_\phi~ \hat{\phi} \nonumber \\
&=& J_\eta \hat{\eta} +J_\theta \hat{\theta}+J_\phi \hat{\phi},
\label{equation1}
\end{eqnarray}
and using the Dirac brackets they satisfy the algebra

\begin{equation}
\{J_\eta,J_\theta\}=-\frac{\alpha\cosh\eta_0\sinh\eta_0}{\alpha^2+\sinh^2\eta_0}p_\phi,~\{J_\phi,J_\eta\}=-\frac{\alpha^2\cosh\eta_0}{\alpha^2+\sinh^2\eta_0}p_\phi,~\{J_\theta,J_\phi\}=0.
\end{equation}
The coordinate transformation rule now changes to
\begin{eqnarray}
\{\vec{J},\phi\} =\frac{\cosh\eta_0\sinh\eta_0}{(\alpha^2+\sinh^2\eta_0)}\sin\alpha\phi~\hat{\eta}-\frac{\sinh^2\eta_0}{(\alpha^2+\sinh^2\eta_0)}\cos\alpha\phi~\hat{\theta} +\frac{\alpha\sinh\eta_0}{(\alpha^2+\sinh^2\eta_0)}\cos\alpha\phi~\hat{\phi}.
\label{jphi}
\end{eqnarray}
It is interesting to note that although the constrained system has been reduced to a single variable one ie. $\phi $,  due to the twisted nature of the particle orbit there are two non-trivial operators $J_\eta,J_\theta $ (but $J_\phi $ and $J_\theta $ are not independent). For this reason $\{\vec{J},\theta\}$ is identical to (\ref{jphi}) apart from a scaling by $\alpha $. Also note that though the angles $\theta,~\phi$ are related by the constraint $\psi _1$ the unit vectors $ \hat\theta,~\hat\phi$ remain independent.

\subsection{Cartesian coordinate system}
To facilitate a comparison with the particle on a sphere case let us analyze the system in a Cartesian framework where the constraint $\theta-\alpha\phi \approx 0$ becomes,
\beq
\psi_1=\cos^{-1}\left(\frac{r^2-a^2}{\sqrt{\phi}}\right)-\alpha\tan^{-1}\frac{x_2}{x_1},
\label{t13}
\eeq
with  $$\phi=(r^2-a^2)^2+4a^2x_3^2=\frac{4a^2(r^2-x_3^2)}{\sinh^2\eta_0}.$$ The Hamiltonian on the other hand has the simple form,
\begin{eqnarray} H=\frac{p_ip_i}{2m}+\lambda \psi_1 .
\label{t14}
\end{eqnarray}
Taking time derivative of $\psi_1$ generates $\psi_2$,
\beq
\psi_2=\frac{-2ax_3(2x.p)+2a(r^2-a^2)p_3}{\phi}-\frac{\alpha\epsilon_{3lk}x_lp_k}{r^2-x_3^2}.
\label{t15}
\eeq
This leads to the Dirac Brackets (see for example \cite{turski} for similar brackets),
\begin{eqnarray}
\{x_i,x_j\} &=& 0,~\{x_i,p_j\}=\delta_{ij}-\frac{A_iA_j}{A^2}=\delta_{ij}-N_iN_j , \nonumber \\
\{p_i,p_j\}&=& -\frac{p_k}{A^2}\left(A_i\frac{\partial A_j}{\partial x_k}-A_j\frac{\partial A_i}{\partial x_k}\right)=\frac{1}{A^2}\left(A_j(p.\partial)A_i-A_i(p.\partial)A_j\right) \nonumber\\
&=& N_j\left(\textbf{p}.\frac{\partial}{\partial\textbf{x}}\right)N_i-N_i\left(\textbf{p}.\frac{\partial}{\partial\textbf{x}}\right)N_j.
\label{t16}
\end{eqnarray}
where the unit normal $N_i$ to the constraint surface $\psi_1$  stands for
$N_i=\frac{A_i}{|A|},$
\begin{eqnarray}
&& A_i=\left[\frac{-4ax_3x_i+2a(r^2-a^2)\delta_{3i}}{\phi}-\frac{\alpha\epsilon_{3li}x_l}{r^2-x_3^2}\right]  \nonumber \\ && ~~~=-\frac{1}{r^2-x_3^2}\left[\frac{\sinh^2\eta_0}{2a}\left(2x_3x_i-(r^2-a^2)\delta_{3i}\right)+\alpha\epsilon_{3li}x_l\right]
\label{t17}
\end{eqnarray}
with $ A^2=\frac{\alpha^2+\sinh^2\eta_0}{r^2-x^2_3}$.

The Dirac algebra is expressed in terms of the unit normal to the $\psi_1$-constraint surface  $\hat{N}_i$. Comparing with our results (\ref{t7}) of this problem in toroidal coordinates we find that the results in Cartesian coordinates depend both on $a$ and $\alpha $ but in fact using the constraints (which depend on $a$ contrary to the toroidal coordinate case) $a$ can be removed.

 The equations of motion (on the constraint manifold)  are
\begin{eqnarray}
\dot{x_i}=\frac{p_i}{m},~~
\dot{p_i}=-\frac{p_kp_j}{m}\frac{\partial N_j}{\partial x_k} N_i
\label{t19}
\end{eqnarray}
leading to
\beq
\ddot{x_i}=-\frac{p_kp_j}{m^2}\frac{\partial N_j}{\partial x_k} N_i.
\label{t20}
\eeq
In explicit form it reads
\begin{eqnarray}
\ddot{x_i} = \left[\frac{\frac{\sinh^2\eta_0}{a}p^2x_3+\frac{\sinh^2\eta_0}{a^2(r^2-x_3^2)} (r^2+a^2)(p_2x_1-x_2p_1)\alpha p_3x_3}{(r^2-x_3^2)} 
+\frac{2\alpha x_3p_3(x_1p_2-x_2p_1)}{(r^2-x_3^2)^2}\right] \frac{N_i}{m^2A}.
\end{eqnarray}
For the canonical structure of angular momentum $J_i=\epsilon_{ijk}x_jp_k$,
\beq
\dot{J_i}=-\frac{1}{m}\epsilon_{ilk}x_lN_k\left(\frac{\partial N_r}{\partial x_j} p_rp_j\right).
\label{js}
\eeq
The fact that $\dot {J_i}$ is non-vanishing is not surprising since from (\ref{t20}) is no longer radial and infact $\dot {J_i}=(\vec r \times \ddot {\vec x})_i $. However, the canonical form of angular  momentum may not be very useful since it induces non-canonical transformations on $x_i,p_i$
\begin{eqnarray}
\{J_i,x_j\} &=& \epsilon_{ijl}x_l+\epsilon_{ilk}x_lN_kN_j, \label{jx} \\
\{J_i,p_j\} &=& \epsilon_{ijk}p_k+\epsilon_{ilk}\left[-N_lN_jp_k+x_l\left(N_j(p.\partial)N_k-N_k(p.\partial)N_j\right)\right],
\label{jp}
\end{eqnarray}
in such a way that even the norms $r^2$ and $p^2$ are not preserved under rotation. The  $J_i$s satisfy the  algebra,
\begin{eqnarray}
\{J_i,J_j\}&=&(x.N)(N_ip_j-N_jp_i)+(x.N)[x_i(p.\partial)N_j-x_j(p.\partial)N_i] \nonumber \\ &&-r^2[N_i(p.\partial)N_j-N_j(p.\partial)N_i]+\left(x_k(p.\partial)N_k\right)(x_jN_i-x_iN_j).
\label{jj}
\end{eqnarray}
Generically in situations like this one tries to construct Darboux type of degrees of freedom that transform canonically (at least approximately if not exactly) which we have postponed for a future work. 

It should be pointed out that the present analysis is not complete and a differential geometric analysis, considering the particle on torus knot trajectory as a space curve, might be more appropriate. It is straightforward to construct a local orthogonal coordinate system known as  triad (or trihedral) consisting of the tangent, principal normal and bi-normal for the space curve. Subsequently, as the particle moves, the moving trihedral is described by a coupled set of differential equations, known as Frenet-Serret relations. As a future work our aim is to generalize the above system to a spinning particle on torus knot and the intrinsic spin vector can be expressed in the trihedral and its interaction with external fields can be studied by exploiting the Frenet-Serret relations.

\section{ Charged particle on a torus knot in external magnetic field: Semi-classical aspects}
In this section we will consider non-trivial effects of the restricted configuration space in particle dynamics. Our approach will be semi-classical quantization to compute particle energy spectrum.

We will use a simpler coordinate system since we are focusing our attention on a particular torus so that there are only two degrees of freedom $\theta $ and $\phi $. The parametric equation for a torus is given by,
\begin{eqnarray}
x_1=(a\sin\theta+d)\cos\phi,~x_2=(a\sin\theta+d)\sin\phi,~x_3=a\cos\theta
\label{5}
\end{eqnarray}
where $\theta$ and $\phi$ are in interval $[0,2\pi]$ , $d$ is the distance from the center of the tube to the center of the torus axis and $a$ is the radius of the tube (see Figures 1 and 2). The equation  for a torus symmetric about the $x_3$-axis is given by
\begin{eqnarray}
(\sqrt{x_1^2+x_2^2}-d)^2+x_3^2=a^2.
\label{5a}
\end{eqnarray}
Exploiting the identities,
\begin{eqnarray}
\dot{x_1}^2+\dot{x_2}^2+\dot{x_3}^2=a^2\dot{\theta}^2+(a\sin\theta+d)^2\dot{\phi}^2, & x_1\dot{x_2}-x_2\dot{x_1}=(a\sin\theta+d)^2\dot{\phi}],
\label{6}
\end{eqnarray}
the Lagrangian for a charge moving on a toroidal surface in a uniform $B$ along $x_3$ is given by,
\begin{eqnarray}
L=\frac{m}{2}[a^2\dot{\theta}^2+(a\sin\theta+d)^2\dot{\phi}^2]-\frac{eB}{2}(a\sin\theta+d)^2\dot{\phi}.
\label{7}
\end{eqnarray}
It should be noted that we have considered a magnetic field which, in technical terms, is referred as solenoidal magnetic field. For a particle on a torus knot, two other forms of magnetic field, referred to as toroidal  (along angle $\phi$ in Figure 1) and poloidal  (along angle $\theta$ in Figure 1) forms are also relevant and should be included in a more general setup.
\begin{figure}[htb!]
	{\centerline{\includegraphics[width=9cm, height=6cm] {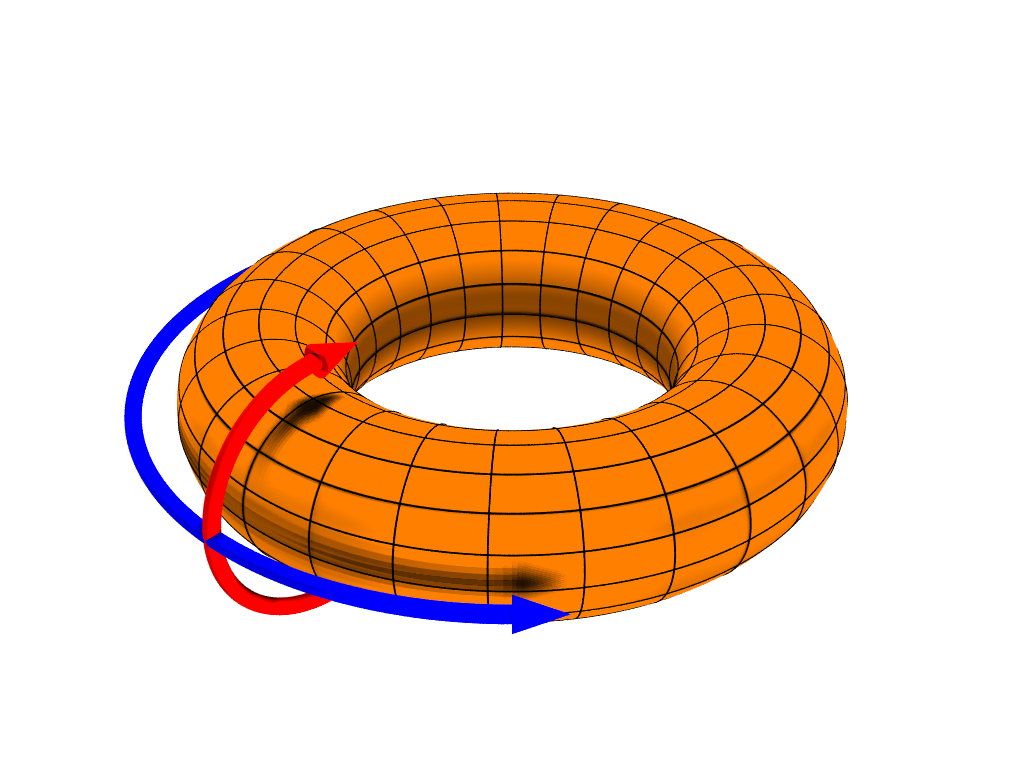}}}
	\caption{Toroidal surface with blue and red lines showing the directions of $\phi $ and $\theta $ respectively.} \label{fig1}
\end{figure}
\begin{figure}[htb!]
	{\centerline{\includegraphics[width=10cm, height=6cm] {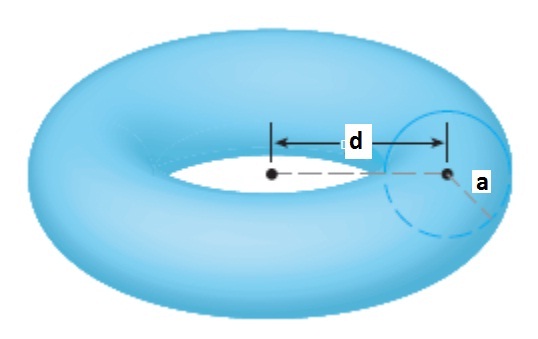}}}
	\caption{The parameters $d$ and $a$ for a toroidal surface are shown.} \label{fig2}
\end{figure}
\begin{figure}[htb!]
	{\centerline{\includegraphics[width=9cm, height=6.5cm] {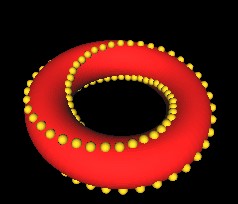}}}
	\caption{The $(2,3)$ trefoil torus knot.} \label{fig3}
\end{figure}

From (\ref{7}), the conjugate momenta are,
\begin{eqnarray}
p_\theta=ma^2\dot{\theta} \\
p_\phi=(m\dot{\phi}-\frac{eB}{2})(a\sin\theta+d)^2=c_1,
\label{p4}
\end{eqnarray}
with $c_1$ a constant since $\phi $ is a cyclic coordinate. The Hamiltonian takes the form,
\begin{eqnarray}
H=\frac{p_\theta^2}{2ma^2}+\frac{[p_\phi+\frac{eB}{2}(a\sin\theta+d)^2]^2}{2m(a\sin\theta+d)^2}.
\label{h}
\end{eqnarray}	

The equations of motion for $\theta$ and $\phi$  are,
\begin{eqnarray}
ma\ddot{\theta}=(a\sin\theta+d)(m\dot{\phi}^2-eB\dot{\phi})\cos\theta ,
\label{lzz}
\end{eqnarray}
\begin{eqnarray}
(m\dot{\phi}-\frac{eB}{2})(a\sin\theta+d)^2=c_1   \end{eqnarray}
or equivalently,
\begin{equation}
m\dot{\phi}=\frac{c_1}{(a\sin\theta+d)^2}+\frac{eB}{2}.
\label{01}
\end{equation}
From the $\ddot \theta $-equation a first integral is easily obtained, 
\begin{eqnarray}
\dot{\theta}^2=-\frac{c_1^2}{m^2a^2(a\sin\theta+d)^2}-\frac{e^2B^2}{4m^2a^2}(a\sin\theta+d)^2+c_2
\label{p5}
\end{eqnarray}
where $c_2$ is another constant.	

The semi-classical quantization is very closely related to the  Hamilton-Jacobi formalism. The idea is to look for a canonical transformation to a new set of variables - action ($J$) angle variables - such that $J$ is cyclic and conserved. The characteristic function $W$ is a solution of the  Hamiltonian-Jacobi equation,
\begin{equation}
H(q_i , \partial W/\partial q_i )-E=0.
\label{hj}
\end{equation}
The action variables are
\begin{equation}
J_i=\oint p_i dq_i = \oint \frac{\partial W}{\partial q_i} dq_i  .
\label{12}
\end{equation}
The quantization conditions, imposed on $J_i$, the adiabetic invariants,  for the conjugate pair  $q_i, p_i$ is given by,
\begin{eqnarray} \label{b1}
\frac{1}{2\pi}\oint p_i dq_i=(n_i+\frac{\mu _i}{4})\hbar ,
\end{eqnarray}
where where $n_i=0,1,2,..$ and $\mu _i$ is the Maslov index \cite{mas} (for applications see for example \cite{est}). It denotes the total phase loss during one period and contributes $1$ or $2$ unit depending on soft (vanishing momentum) or hard (reflection) classical turning points respectively. In the present case we will have two quantization conditions
\begin{eqnarray}\label{q1}
\frac{1}{2\pi}\oint p_\phi d\phi=n_\phi\hbar ,
\end{eqnarray}
\begin{eqnarray}\label{q2}
\frac{1}{2\pi}\oint p_\theta d\theta =n_\theta\hbar .
\end{eqnarray}
Maslov index does not contribute since both $\theta$ and $ \phi $ are rotational degrees of freedom (and not librations). The first one, (\ref{q1}) is trivial and produces,
\begin{eqnarray}\label{lz}
p_\phi=n_\phi\hbar.
\end{eqnarray}
However, the second cyclic integral (\ref{q2}) is non-trivial which we perform in a slightly unconventional way, as explained in Appendix III. The reason is that, in case of toroidal coordinates as has been used here, the integral is more complicated than the well known one in spherical polar coordinates. In the Appendix III we have discussed the known case of  spherical coordinate within our formalism and have exploited the same technique for the toroidal case without any ambiguity.

The phase integral 
\begin{eqnarray}
	\oint p_\theta d\theta&=&\oint ma^2\dot{\theta}d\theta \nonumber \\
	&=&\oint ma^2\dot{\theta}^2 dt  \nonumber \\
	&=& ma^2\oint \left[c_2-\frac{c_1^2}{m^2a^2(a\sin\theta+d)^2}-\frac{e^2B^2}{4m^2a^2}(a\sin\theta+d)^2\right]dt \nonumber \\
	&=& ma^2\oint \left[c_2-\frac{c_1}{m^2a^2}(m\dot{\phi}-\frac{eB}{2})-\frac{eB}{2m^2a^2}(m\dot{\phi}(a\sin\theta+d)^2-c_1)\right]dt \nonumber \\
	&=& ma^2\oint\left[(c_2+\frac{eBc_1}{m^2a^2})-\frac{c_1}{ma^2}\dot{\phi}-\frac{eB}{2ma^2}(a\sin\theta+d)^2\dot{\phi}\right] dt 
	\label{p1}
\end{eqnarray}

is rewritten as,
\begin{eqnarray}
\oint p_\theta d\theta&=& ma^2(c_2+\frac{eBc_1}{m^2a^2})\oint dt-\oint p_\phi d\phi-\frac{eB}{2}\oint (a\sin\theta+d)^2 d\phi \nonumber \\
&=& ma^2(c_2+\frac{eBc_1}{m^2a^2})\frac{l}{v}-2\pi p_\phi-\frac{eB}{2}2\pi p (\frac{a^2}{2}+d^2)
\label{a1}
\end{eqnarray}
using time period $T=\frac{l}{v}$. In the above, we have imposed the knot condition  $\theta=q\theta$ and $\phi=p\theta$ and  $l=\int_{0}^{2\pi}\sqrt{q^2a^2+p^2(d+a\sin q\theta)^2}d\theta$ constitutes the arc length for the trajectory and $v$ is the velocity {\footnote{We have used the $(p,q)$-torus knot parameterization $x_1=(a\sin(q\theta )+d)\cos(p\theta ),~x_2=(a\sin(q\theta )+d)\sin(p\theta ),~x_3=a\cos(q\theta ),~0\le\theta \le 2\pi  $}}.

From the definition of velocity in toroidal coordinates,  it follows that  the velocity is 
\begin{eqnarray}
v^2=a^2 \dot{\theta}^2+(a\sin\theta+d)^2\dot{\phi}^2
=a^2c_2+\frac{eBc_1}{m^2}.
\label{a2}
\end{eqnarray}
Hence, the time period $T=\oint dt$ is given by 
\begin{eqnarray}
T=\frac{l}{v}=\frac{\int_{0}^{2\pi}\sqrt{q^2a^2+p^2(d+a\sin q\theta)^2}d\theta}{\sqrt{a^2c_2+\frac{eBc_1}{m^2}}}
\label{p2}
\end{eqnarray}

The trajectory length $l$ 
\begin{eqnarray}
l&=&\int_{0}^{2\pi} \sqrt{q^2a^2+p^2(d+a\sin q\theta)^2} d\theta \nonumber \\
&=& pd\int_{0}^{2\pi} \sqrt{1+\frac{2a}{d}\sin q\theta+\frac{a^2}{d^2}(\frac{q^2}{p^2}+\sin^2q\theta)}  d\theta .
\label{p7}
\end{eqnarray}
for $d>>a$, in a thin torus approximation, to  order  $O(\frac{a^2}{d^2})$,  reduces to,
\begin{eqnarray}
l\approx 2\pi pd \left(1+\frac{a^2q^2}{2p^2d^2}\right).
\label{p8}
\end{eqnarray}

Thus the quantization condition yields
\begin{eqnarray}
\oint p_\theta d\theta = ml\sqrt{(a^2c_2+\frac{eBc_1}{m^2})}-2\pi p_\phi-\frac{eB}{2}2\pi p (\frac{a^2}{2}+d^2) = 2\pi n_\theta\hbar.
\label{a3}
\end{eqnarray}

\subsection { Discrete energy spectrum}
Let us now focus on the semi-classical quantization problem. The  Schrodinger or wave equation is not separable in toroidal coordinate system. In the semi-classical scheme we exploit the  Einstein - Brillouin - Keller (EBK) \cite{ebk} scheme of  quantization appropriate for  multidimensional
systems (for applications see for example \cite{ebk1}). Note that although the system, after imposing the knot condition essentially reduces to a one dimensional system, even then it has manifest three dimensional features. Interestingly this again shows up later in Section 3.2 when we study the fractional angular momentum.   In multidimensional systems the periodic motion of the particle  is restricted to a set of invariant torus in phase space and EBK formalism imposes a quantization condition for a path integral in phase space for each coordinate and its conjugate momentum.

 In fact we have already provided the quantization conditions in (\ref{lz},\ref{a3}). What remains is simply to replace the constants $c_1,c_2$ using the relations (\ref{lzz},\ref{p5}) in (\ref{h}), expression for the energy.

  From the Hamiltonian (\ref{h}) we can write,
  \begin{eqnarray}
  \oint p_\theta d\theta=\oint\sqrt{2mEa^2-\frac{p_\phi^2a^2}{(a\sin\theta+d)^2}-p_\phi eB a^2-\frac{e^2B^2a^2}{4}(a\sin\theta+d)^2}   d\theta ,
  \label{p12}
  \end{eqnarray}
   and from (\ref{p5}) we find,
  \begin{eqnarray}
  \oint p_\theta d\theta=\oint\sqrt{m^2a^4c_2-\frac{p_\phi^2a^2}{(a\sin\theta+d)^2}-\frac{e^2B^2a^2}{4}(a\sin\theta+d)^2} d\theta .
  \label{p13}
  \end{eqnarray}
  Comparing the two relations above we see that
  \begin{eqnarray}
  c_2=\frac{2mE-p_\phi eB}{m^2a^2}.
  \label{p14}
  \end{eqnarray}

  After some straightforward algebra we obtain the cherished energy spectrum, to  order  $O(\frac{a^2}{d^2})$, 
 \begin{eqnarray}
 E_n=\frac{1}{2m}\left[\frac{n^2\hbar^2}{p^2d^2} \left(1-\frac{a^2q^2}{p^2d^2} \right)+\frac{eB}{p}n\hbar \left(1+\frac{a^2}{2d^2}-\frac{a^2q^2}{p^2d^2}\right)+\frac{e^2B^2d^2}{4} \left(1+\frac{a^2}{d^2}-\frac{a^2q^2}{p^2d^2}\right)\right],
 \label{p11}
 \end{eqnarray}
 where $n=n_\theta + n_\phi $.

\textbf{Effective one dimensional model:} Once again it is possible to do a quick computation in Bohr-Sommerfeld framework, treating the system as effectively one dimensional once the torus knot condition is imposed. It will be interesting to compare the result with  more elaborate computation done earlier. The single variable Lagrangian of a charge, moving along a torus knot,  in the presence of  $B$ along the symmetry axis of the torus, is obtained from (\ref{7}) as,
\begin{eqnarray}
L=\frac{m}{2}[a^2q^2\dot{\theta}^2+(a\sin[q\theta]+d)^2p^2\dot{\theta}^2]-\frac{eBp}{2}(a\sin[q\theta]+d)^2\dot{\theta}.
\label{8}
\end{eqnarray}
The conjugate momentum and Hamiltonian are respectively given by, 
\begin{eqnarray}
p_\theta=m[a^2q^2+(a\sin[q\theta]+d)^2p^2]\dot{\theta}-\frac{eBp}{2}(a\sin[q\theta]+d)^2,
\label{9}
\end{eqnarray}
\begin{eqnarray}
H=E=\frac{[p_\theta+\frac{eBp}{2}(a\sin[q\theta] +d)^2]^2}{2m[a^2q^2+p^2(a\sin[q\theta]+d)^2]}.
\label{10}
\end{eqnarray}\\
In the present case we find,
\begin{eqnarray}
p_\theta=\frac{\partial W}{\partial \theta}=\sqrt{2mE[a^2q^2+p^2(a\sin[q\theta]+d)^2]}-\frac{eBp}{2}(a \sin[q\theta] +d)^2
\label{11}
\end{eqnarray}
 For $a=0$ the toroidal coordinate system (\ref{5},\ref{5a}) reduces to a circle of radius $d$ on the $x_1-x_2$ plane. We restrict ourselves to the thin torus limit,  $d>>a$. Notice that the    $O(a/d)$ correction terms do not contribute in the integral (\ref{b1}) and so we consider  results up to the first non-trivial order that is $O(a/d)^2$. We find, to order of $O(a^2/d^2)$, 
\begin{eqnarray}
p_\theta = p d \sqrt{2mE} \left[ 1 + \frac{1}{2} \left(\frac{a^2q^2}{d^2p^2} + \frac{2a\sin[q\theta]}{d} \right) \right] - \frac{eBpd^2}{2} \left[ 1 + \frac{a^2\sin^2[q\theta]}{d^2} + \frac{2a\sin[q\theta]}{d}\right] \label{en}
\end{eqnarray}	
To invoke the quantization condition (\ref{b1}) we need to fix the limits of  integration of $\theta $. Observe that  $a=0$   will yield the correct expressions for a charge moving in a ring of radius $d$ with $B$ normal to the plane of motion the limits of $\theta $ has to be  $\theta =0 $ to  $\theta = 2\pi $. It should be pointed out that the factor $p$ appears since the charge makes $p$ revolutions with radius $d$ in the limit $a=0$. For a circle $p$ is unimportant and is fixed to $p=1$ but for the torus knot both $p,q$ specifies the path and so are kept arbitrary. The energy spectrum derived is identical to the earlier one (\ref{p11}) but indeed, now with the single quantum number $n$. This indicates that particle in a torus knot has same type of degeneracy in energy spectrum as the motion in a central force.

 The energy levels for the particle on a circle with $B=0$, are given by
\begin{equation}
E_n= \frac{(n \hbar)^2}{2md^2}.
\label{4a}
\end{equation}
This can be compared with the case of particle on a torus knot also with $B=0$,
\begin{equation}
E_n=\frac{(n\hbar )^2}{2md^2p^2}\left [ 1- \frac{q^2}{p^2}\frac{a^2}{d^2}\right].
\label{4aa}
\end{equation}
Clearly the latter contains more structure due to the  complicated nature of the path in the form of a torus knot. The expressions match for $a=0$ and $p=1$.\\

\subsection{Fractional angular momentum}

To see the effect of the knot on the fractional angular momentum let us reexpress (\ref{9}) as
\begin{eqnarray}
p_\theta&=&m[a^2q^2+(a\sin[q\theta]+d)^2p^2]\dot{\theta}-\frac{eBp}{2}(a\sin[q\theta]+d)^2 \nonumber \\
&=& L-\frac{eBp}{2}(d^2+a^2\sin^2q\theta+ 2 a d \sin[q\theta]),
\label{am}
\end{eqnarray}
where $L$ stands for the kinetic angular momentum in absence of $B$. It needs to be mentioned that $L$ is not conserved since the path is no longer restricted to a plane. Integrating the $B$-dependent term along the closed path yields 
\begin{equation}
\oint \left[\frac{eBp}{2}(d^2+a^2\sin^2[q\theta] + 2 a d \sin[q\theta]) \right] d\theta =\frac{eBp}{2} \times 2\pi \left(d^2+\frac{a^2}{2} \right).
\label{}
\end{equation}
Therefore, on imposition of the quantization condition
\begin{eqnarray}
\oint p_\theta d\theta=nh,
\label{qq}
\end{eqnarray}
we find that although the total angular momentum changes by discrete steps the absolute value is in general non-integral due to the additional factor  of $\frac{eBp}{2}2\pi(d^2+\frac{a^2}{2})$. Once again this will match with the circular path for $a=0, p=1$. Note that it explicitly depends on both the torus knot parameters $(d,a)$, specifying the geometry of the embedding torus, and $(p,q)$, the parameters specifying the knot. Surprisingly,  the result, at least to $O(a/d)^2$,  is independent of $q$. Dependence of the fractional spin on both the torus knot parameters $q,p$ where $q$ in particular signifies the non-planar feature of the path justifies our claim that non-planar paths can affect the anyon properties \cite{wil,others}.\\

\section {Conclusion and future prospects}
Let us summarize the work reported in this paper. Our main objective is to study the motion of a particle on a torus knot in Hamiltonian framework. The system has a rich constraint structure and Dirac's theory of constraint dynamics is exploited to analyze the dynamics and kinematics in detail. The system consists of Second Class constraints in Dirac terminology and computation of the Dirac brackets yields a novel form of noncommutative (or non-canonical) algebra whose  commutative limit is subtle since it does not simply depend on (numerical) noncommutative parameters, as  is the case with other noncommutative algebra found in the literature.

We have constructed the particle motion in a generalized set up where the particle is constrained to execute the knot but the motion is not restricted to any particular torus. This means that in the conventional toroidal coordinate system, the coordinates $\theta $ and $\phi $ are identified via the constraint (that imposes the knot in motion) but the coordinate $\eta $ that fixes a specific torus is allowed to be dynamical. This indicates that fluctuations of the particle trajectory are considered keeping the non-trivial topology of the path (torus knot) intact.  Later on we further restrict the motion by constraining $\eta$ to a fixed value.

In the context of kinematics we have studied the nature of the angular momentum operator and have shown that an improved form of it is required for a consistent analysis of the problem. Lastly we have analyzed the behavior of small fluctuations in the particle motion  about the torus knot.

In the second part of the  paper we have considered the quantum theory of a charge moving in a torus knot in the presence of a uniform magnetic field along the axis of the torus in a semiclassical quantization framework. We exploit the  Einstein - Brillouin - Keller (EBK)  scheme of  quantization that is appropriate for  multidimensional
systems. Note that although the system, after imposing the knot condition essentially reduces to a one dimensional system, even then it has manifest three dimensional features.  Interestingly this again shows up later in the study of fractional angular momentum.  We show that the energy levels and fractional spin depend on the torus knot parameters that specifies its non-planar features. Finally we compare the results obtained from  EBK (multi-dimensional) and  Bohr-Sommerfeld (single dimensional) scheme.

Let us now elaborate on the possible extensions of the present work and open problems. Concerning the formal constraint dynamics perspective,  construction of Darboux-like canonical degrees of freedom will be worthwhile as it will pave the way for quantization of the generalized system. Also the path of the particle with the torus knot can lead to non-trivial homotopy features that will be manifest in semi-classical quantization conditions. 

In the generalization of the particle model interacting with the external magnetic field, it will be interesting to attribute an intrinsic spin degree of freedom to the charged particle since then the topological nature of the (torus) knotted path should become more manifest.

Another possible extension is to consider more general forms of interaction, (such as different forms of magnetic field), so that the degeneracy in energy spectrum can belifted.

Finally notice that there is a fundamental distinction between our approaches: in the first part of the paper we worked in Cartesian coordinates where the constraints on the particle motion were manifest and subsequently the Poisson brackets were replaced by Dirac brackets. But in the second part where we treated semi-classical quantization we exploited suitably reduced coordinate systems where Dirac brackets were not necessary. Indeed it would be interesting if we can derive the semi-classical quantization results in Cartesian coordinate framework with non-canonical Dirac bracket algebra. This is possible provided a generalization of the semi-classical quantization prescription is adopted. In a recent work \cite{mig} this approach has been used to compute energy spectra that is compatible to a different form of Dirac brackets.\\

\paragraph{Appendix I:}
In the terminology of Dirac constraint analysis \cite{dir}, the
noncommutating constraints are termed as SCC and the commutating constraints, that induces local gauge invariance, are named First Class Constraints (FCC). In a generic Second Class system with $n$  SCCs $\chi_i$, $i=1,2,..n$, the modified symplectic structure (or Dirac brackets) are defined in the following way,
\begin{equation}
\{A,B\}^*=\{A,B\}-\{A,\chi _i\}\{\chi ^i,\chi ^j\}^{-1}\{\chi _j,B\}, \label{a6}
\end{equation}
where $\{\chi ^i,\chi ^j\}$ is the invertible constraint matrix. From now on we will use $\{,\}$ notation instead of $\{,\}^*$ for Dirac brackets.

In Cartesian coordinate system, the  non-zero constraints matrix element reads
\begin{equation}
\{\psi_1(r),\psi_2(p,r)\}=\{\psi_1(r),p.A(r)\}=\frac{\partial \psi_1}{\partial x_i}A_i=A_iA_i=A^2
\end{equation}
where, $A_i=\frac{\partial \psi_1}{\partial x_i},~A^2=\frac{4a^2}{\phi}+\frac{\alpha^2}{r^2-x_3^2}.$
Thus the inverse matrix element can be written as,
\begin{equation}
\{\psi_1(r),\psi_2(p,r)\}^{-1}=-\frac{1}{A^2}
\end{equation}

The Dirac bracket can be computed in the following way,
\begin{eqnarray}
\{x_i,p_j\}_{D.B.}=\delta_{ij}-\{x_i,\psi_2\}\{\psi_2,\psi_1\}^{-1}\{\psi_1,p_j\}
=\delta_{ij}-\{x_i,p.A\}\{\psi_2,\psi_1\}^{-1}\frac{\partial \psi_1}{\partial x_j}=\delta_{ij}-\frac{A_iA_j}{A^2}.
\end{eqnarray}

\paragraph{Appendix II:}
In toroidal coordinate system, the  non-zero constraints matrix element is
$$\{\chi_1,\chi_2\}=\frac{p^2 \sinh^2\eta+q^2}{\sinh^2\eta}.$$
Thus the inverse matrix element can be written as
\begin{equation}
\{\chi_1,\chi_2\}^{-1}=-\frac{\sinh^2\eta}{p^2\sinh^2\eta+q^2}.
\end{equation}

The Dirac bracket $$\{\phi,p_\phi\}$$ can be computed in the following way,
\begin{equation}
\{\phi,p_\phi\}_{D.B.}=\{\phi,p_\phi\}-\{\phi,\chi_2\}\{\chi_2,\chi_1\}^{-1}\{\chi_1,p_\phi\}$$$$
=1-\frac{q^2}{p^2\sinh^2\eta+q^2}=\frac{\sinh^2\eta}{\alpha^2+\sinh^2\eta}
\end{equation}
where $\alpha=-\frac{q}{p}$.

\paragraph{Appendix III:}
In spherical coordinates, the Lagrangian for a  particle to move on a sphere of radius $a$,
\begin{eqnarray}
L=\frac{m}{2}(a^2\dot{\theta}^2+a^2\sin^2\theta\dot{\phi}^2)-V
\label{a6}
\end{eqnarray}
gives rise to the Hamiltonian, 
\begin{eqnarray}
H=\frac{1}{2ma^2}(p_\theta^2+\frac{p_\phi^2}{\sin^2\theta})+V
\label{a7}
\end{eqnarray}
and from the equations of motion we recover the integrals of motion,
\begin{eqnarray}
ma^2\sin^2\theta\dot{\phi}=p_\phi =c_1,~~ \dot{\theta}^2=c_2-\frac{c_1^2}{m^2a^4\sin^2\theta}
\label{a8}
\end{eqnarray}
with $c_1,c_2$ being constants. Furthermore we find
\begin{eqnarray}
p_\theta^2+\frac{p_\phi^2}{\sin^2\theta}=c_2m^2a^4.
\label{a9}
\end{eqnarray}
and the velocity
\begin{eqnarray}
v^2=a^2\dot{\theta}^2+a^2\sin^2\theta\dot{\phi}^2= a^2c_2 ~~\rightarrow v=a\sqrt{c_2}.
\label{a10}
\end{eqnarray}
To impose the $\theta $-quantization condition we need to compute the following integral:
\begin{eqnarray}
\frac{1}{2\pi}\oint p_\theta d\theta&=&\frac{1}{2\pi}\oint ma^2\dot \theta d\theta =   \frac{1}{2\pi}\oint ma^2\dot{\theta}^2dt \nonumber \\
&=& \frac{1}{2\pi}\oint ma^2[c_2-\frac{c_1^2}{m^2a^4\sin^2\theta}]dt \nonumber \\
&=& \frac{ma^2 c_2}{2\pi} \oint dt- \frac{c_1}{2\pi}\oint \dot{\phi} dt \nonumber \\
&=& \frac{ma^2 c_2}{2\pi} \frac{2\pi a}{a\sqrt{c_2}}- \frac{c_1}{2\pi} 2\pi \nonumber \\
&=& L-L_z .
\label{a11}
\end{eqnarray}In the last but one step we note that the time period $\oint dt =T=(2\pi a)/v =(2\pi a)/(a\sqrt{c_2})$ and to match with the notation of Goldstein \cite{gold} we identify 
$$p_\phi =c_1\equiv L_z,~p_\theta^2+\frac{p_\phi^2}{\sin^2\theta}=c_2m^2a^4\equiv L^2,$$ which reproduces the correct result. Indeed it is possible to obtain this result directly by performing the $\theta $-integral \cite{gold} but in the toroidal coordinate that is of present interest the $\theta $-integral turns out to be more complicated and so we use the same technique in the toroidal case in the main text.

\vskip .5cm
{\bf{Acknowledgement:}}
P.D. acknowledges the financial support from INSPIRE, DST, India. We thank the referee for the constructive comments that have helped us to improve presentation of the paper.

\end{document}